\documentstyle[11pt,newpasp]{article}

\begin{document}

\title{Phase--Space Constraints on Visible and Dark Matter Distributions
       in Elliptical Galaxies}

\author{L. Ciotti}
\affil{Osservatorio Astronomico di Bologna, via Ranzani 1, I-40127 Bologna}
\affil{Scuola Normale Superiore, Piazza dei Cavalieri 7, I-56126 Pisa}

% The abstract is entered in a LaTeX "environment", designated with paired
% \begin{abstract} -- \end{abstract} commands.  Other environments are
% identified by the name in the curly braces.

% Poster authors ONLY may omit the abstract in order to gain a little
% more page space for the text of the poster.

\begin{abstract}
There are observational and theoretical indications that both the
visible (stars) and the dark matter density distributions in
elliptical galaxies increase significantly up to the galactic
center. I present here some analytical results obtained with the aid
of self--consistent, spherically symmetric two component galaxy
models. These results suggest the possibility that this similar
behavior could be a direct consequence of the structural and dynamical
constraints imposed by the request of positivity of the phase--space
distribution function (DF) of each density component.
\end{abstract}

% Keywords should be included, but they are not printed in the hardcopy.

\keywords{Galaxies: elliptical and lenticulars, kinematics and dynamics,
          photometry}

% That's it for the front matter.  On to the main body of the paper.
% We'll only put in tutorial remarks at the beginning of each section
% so you can see entire sections together.

\section{Introduction}

High resolution observations of elliptical galaxies and bulges of
spirals show that their (spatial) central stellar densities
are well described by a power--law profile, 
\begin{equation}
\rho_*(r)\propto r^{-\gamma},\quad r\sim 0,
\end{equation} 
where $0\leq\gamma <2.4$ (e.g., Gebhardt et al. 1996; Carollo \&
Stiavelli 1998).  Moreover, it is now commonly accepted that a non
negligible fraction of the total mass in elliptical galaxies is made
of a dark component, whose density distribution at large radii differs
significantly from that of the visible one. The radial distribution of
the dark matter in the galactic central regions is less well known. In
the past a commonly used, centrally flat model for the dark matter
distribution was the so called quasi isothermal (QI) density
distribution ($\beta=2$ in eq. [6]), but, more recently, high
resolution numerical simulations of dark matter halo formation showed
that also for the dark matter density distribution
\begin{equation}
\rho_{\rm DM}(r)\propto r^{-\beta}
\end{equation}
in the central regions, where $\beta\geq 1$ (e.g., Dubinsky \&
Carlberg 1991; Navarro, Frenk, \& White 1997). Thus, {\it there are
indications that both the visible and the dark matter density
distributions increase significantly up to the center of galaxies}.  A
natural question arises: is the origin of this qualitatively similar
behavior independent for the two density distributions, or it is
related for some reason?

From a theoretical point of view it is well known that in any
physically acceptable multicomponent galaxy model the phase--space DF
of {\it each} density component must be non negative. A model
satisfying this minimal requirement is called a {\it consistent}
model.  Thus, the question above can be reformulated as follows: {\it
can the request of positivity of the DF of each density component in
multicomponent galaxy models tell us something about the relative
distribution of dark and visible matter in real galaxies?}  In order
to answer this question I studied the problem of the construction and
investigation of the DFs of two component galaxy models with variable
amount and distribution of dark matter, and variable orbital
anisotropy.

\section{Technique}

For a multi component spherical system, where the (radial) orbital
anisotropy of each component is modeled according to the OM
parameterization (Osipkov 1979; Merritt 1985), the DF of the density
component $\rho_k$ is given by:
\begin{equation}
f_k(Q_k)={1\over\sqrt{8}\pi^2}{d\over dQ_k}\int_0^{Q_k} 
{d\varrho_k\over d\Psi_T}{d\Psi_T\over\sqrt{Q_k-\Psi_T}},
\quad
\varrho_k (r)=\left (1+{r^2\over r_{ak}^2}\right) \rho_k (r),
\end{equation}
where $\Psi_T (r)=\sum_{\rm k}\Psi_k (r)$ is the total relative
potential, and $Q_k ={\cal E}-L^2/2r_{ak}^2$.  ${\cal E}$ and $L$ are
respectively the relative energy and the angular momentum modulus per
unit mass, $r_{ak}$ is the {\it anisotropy radius}, and $f_k(Q_k)=0$
for $Q_k\leq 0$ (e.g., Binney \& Tremaine 1987).

Preliminary information on the model consistency can be easily
obtained using the following results (Ciotti \& Pellegrini 1992, CP;
Ciotti 1996, C96; Ciotti 1999, C99).  A {\it necessary} condition for
the non negativity of each $f_k$ is:
\begin{equation}
{d\varrho_k(r)\over dr}\leq 0,\quad 0\leq r \leq\infty .
\end{equation} 
If the NC is satisfied, {\it strong} and {\it weak sufficient}
conditions for the non negativity of each $f_k$ are:
\begin{equation}
{d\over dr}\left[{d\varrho_k(r) \over dr}
{r^2\sqrt {\Psi_T(r)}\over M_T(r)}\right]\geq 0,
\quad
{d\over dr}\left[
{d\varrho_k(r) \over dr}{r^2\over M_T(r)}\right]\geq 0,
\quad 0\leq r\leq\infty .
\end{equation}

The explored two component galaxy models are made of various
combinations of spherically symmetric density distributions, as the
{\it flat--core} mass distributions
\begin{equation}
\rho (r)={\rho_0 r_{\rm c}^{\beta}\over (r_{\rm c}^2+r^2)^{\beta /2}},
\end{equation}
the {\it centrally--peaked} spatial density associated to the
deprojection of the Sersic (1968) profile
\begin{equation}
I(R)=I(0)\exp [-b(m)(R/R_{\rm e})^{1/m}],\quad\rho (r)\sim r^{(1-m)/m},
                                         \quad r\sim 0,
\end{equation}
(where the explicit expression for $b(m)$ can be found in Ciotti \&
Bertin 1999), and finally the $\gamma$ models (Dehnen 1993)
\begin{equation}
\rho (r)={3-\gamma\over 4\pi}
         {M r_{\rm c}\over 
         r^{\gamma}(r_{\rm c}+r)^{4-\gamma}},\quad 0\leq\gamma <3.
\end{equation}
The mass, scale--length, and anisotropy radius of each component are
free parameters. For example, in C96 and C99 the family of two
component self--consistent galaxy models, where one density
distribution follows a $\gamma_1$ profile, and the other a $\gamma_2$
profile [$(\gamma_1,\gamma_2)$ models], is presented.

\section{Results}

The main results can be summarized as follows.

In CP, by {\it numerical} investigation of the inequalities given in
eqs. (4)-(5), it was shown that it is not possible to add an isotropic
QI or Hubble modified halo ($\beta =3$ in eq. [6]) to a R$^{1/4}$
galaxy ($m=4$ in eq. [7], de Vaucouleurs 1948), because their DFs run
into negative values near the model center.  On the contrary, the
isotropic R$^{1/4}$ galaxy was found to be consistent for any value of
the superimposed halo mass and scale--length (i.e., over all the
parameter space).  At variance with the previous case, the isotropic
two component models where both density distributions are
characterized by a centrally flat profile (the Hubble modified + QI
models), or by a centrally peaked profile (the R$^{1/4}$+R$^{1/4}$
models), were found to be consistent over all the parameter space.

In C96 and C99 the analytical DF for both components of OM anisotropic
(1,1) and (1,0) models is found in terms of elliptic
functions. Moreover, the method described by eqs. (4)-(5) is applied
{\it analytically} to the more general family of $(\gamma_1,\gamma_2)$
models. It is proved that for $1\leq\gamma_1 <3$ and
$\gamma_2\leq\gamma_1$ the DF of the $\gamma_1$ component in isotropic
$(\gamma_1,\gamma_2)$ models is nowhere negative, independent of the
mass and concentration of the $\gamma_2$ component.  As a special
application of this result, it follows that a black hole (BH) of any
mass can be consistently added at the center of any isotropic member
of the family of $\gamma$ models when $1\leq\gamma <3$, and that both
components of isotropic $(\gamma,\gamma)$ models (with $\gamma\geq 1$)
are consistent over all the parameter space. As a consequence, the
isotropic $\gamma=1$ component in (1,1) and (1,0) models is
consistent.  In the anisotropic case, it is shown that an analytic
estimate of a minimum value of $r_a$ for one component $\gamma$ models
with a massive BH at their center can be explicitly found. As
expected, this minimum value decreases for increasing $\gamma$. The
region of the parameter space in which (1,0) models are consistent is
successively explored using the derived DFs: it is shown that, unlike
the $\gamma=1$ component, the $\gamma=0$ component becomes
inconsistent when the $\gamma=1$ component is sufficiently
concentrated, even in the isotropic case.  The combined effect of the
mass concentration and orbital anisotropy is also investigated, and an
interesting behavior of the DF of the anisotropic $\gamma=0$ component
is found and explained: it is analytically shown that there exists a
small region in the parameter space where a sufficient amount of
anisotropy can compensate for the inconsistency produced by the
$\gamma=1$ component concentration on the structurally analogous but
isotropic case.

\section{Conclusions}

Some of the general trends that emerge when comparing different one
and two component models with OM anisotropy, as those investigated in
CP, Ciotti, Lanzoni, \& Renzini (1996), Ciotti \& Lanzoni (1997), C96
and C99, can be summarized as follows.

1) In the isotropic case, when the two density components (i.e., stars
and dark matter) are similarly distributed in the galactic center, the
corresponding models are consistent for any choice of the mass ratio
and scale--length ratio.

2) On the contrary, when the two density components are significantly
different in the galactic central regions, the mass component with the
steeper density is ``robust'' against inconsistency. The mass
component with the flatter density in the central regions is the most
``delicate'' and can easily become inconsistent.

3) For sufficiently small values of the anisotropy radius, OM
anisotropy may produce a negative DF outside the galaxy center, while the
halo concentration affects mainly the DF at high (relative) energies
(i.e., near the galaxy center).

4) The trend of the minimum value for the anisotropy radius as a
function of the halo concentration shows that a more diffuse halo
allows for a larger amount of anisotropy: in other words, the
possibility of sustaining a strong degree of anisotropy is weakened by
the presence of a very concentrated halo.

As a consequence of previous points 1 and 2, one could speculate that
in the presence of a centrally peaked dark matter halo, centrally flat
elliptical galaxies should be relatively rare, or that a galaxy with a
central power law density profile cannot have a dark halo too flat in
the center. Thus, {\it the qualitatively similar behavior of visible
and dark matter density profiles in the central regions of elliptical
galaxies could be a consequence of phase--space constraints}.

\acknowledgments

This work was supported by MURST, contract CoFin98.

% That's the end of the main body of the paper.  Now we will have some
% back matter.

% Now comes the reference list.  Since we typed out the citations ourselves,
% the reference list is enclosed in a "references" environment.  Each
% new reference begins with a \reference command which sets up the proper
% indentation.  Typography that may be required in the reference list by
% the editorial staff must be included by the author.
%
% Observe the "standard" order for bibliographic material: author name(s),
% publication year, journal name, volume, and page number for articles.
% Some journal names are available as macros; see the WGAS markup
% instructions for a listing of which ones have been "macro-ized".
% Note the use of curly braces to delimit the font changes: it is essential
% that this be done to limit the scope of the font declaration.
%
% There is no need to engage in any other typographic manipulation.

% That's all, folks.
%
% The technique of segregating major semantic components of the document
% within "environments" is a very good one, but you as an author have to
% come up with a way of making sure each \begin{whatzit} has a corresponding
% \end{whatzit}.  If you miss one, LaTeX will probably complain a great
% deal during the composition of the document.  Occasionally, you get away
% with it right up to the \end{document}, in which case, you will see
% "\begin{whatzit} ended by \end{document}".

\end{document}